\documentclass[twocolumn,aps,prl,superscriptaddress]{revtex4-2}

\usepackage{graphicx}
\usepackage{color}
\usepackage{mathtools}
\usepackage{enumitem}
\usepackage{amssymb}
\usepackage{hyperref}
\usepackage[symbol]{footmisc}

\renewcommand{\vec}[1]{{\bf #1}}

\newenvironment{nalign}{
    \begin{equation}
    \begin{aligned}
}{
    \end{aligned}
    \end{equation}
    \ignorespacesafterend
}

\begin{document}

\title{Tunable electron-phonon interactions in long-period superlattices}

\author{Hiroaki Ishizuka}
\email{hishizuka@g.ecc.u-tokyo.ac.jp}
\affiliation{Department of Physics, Massachusetts Institute of Technology, Cambridge, Massachusetts, 02139, USA}
\affiliation{Department of Applied Physics, The University of Tokyo, Bunkyo, Tokyo, 113-8656, Japan}

\author{Ali Fahimniya}
\affiliation{Department of Physics, Massachusetts Institute of Technology, Cambridge, Massachusetts, 02139, USA}

\author{Francisco Guinea}
\affiliation{Instituto Madrile\~no de Estudios Avanzados en Nanociencia (IMDEA-Nanociencia), 28049 Madrid, Spain}
\affiliation{Donostia International Physics Center (DIPC) UPV/EHU, E-20018, San Sebasti\'an, Spain}

\author{Leonid Levitov}
\affiliation{Department of Physics, Massachusetts Institute of Technology, Cambridge, Massachusetts, 02139, USA}

\date{\today}

\begin{abstract}
The efficiency of optical emitters can be dramatically enhanced by reducing the effective mode volume (the Purcell effect). Here we predict an analogous enhancement for electron-phonon (el-ph) scattering, achieved by compressing the electronic Wannier orbitals. Reshaping of Wannier orbitals is a prominent effect in graphene moir\'e superlattices (SLs) where the orbitals are tunable by the twist angle.  A reduction of the orbital effective volume leads to an enhancement in the effective el-ph coupling strength, yielding the values considerably bigger than those known for pristine monolayer graphene. The enhanced coupling boosts the el-ph scattering rates, pushing them above the values predicted from the enhanced spectral density of electronic excitations. The enhanced phonon emission and scattering rates are manifest in the observables such as electron-lattice cooling and the linear-$T$ resistivity, both of which are directly tunable by the moir\'e twist angle.
\end{abstract}

\maketitle

The discovery of tunable flat-band systems in twisted van der Waals heterostructures~\cite{LopesDosSantos2007,Laissardiere2010,Mele2010,Morell2010,Shallcross2010,Bistritzer2011} has prompted many interesting questions about using the novel twist angle degree of freedom to tune and control the interactions between elementary excitations and the resulting ordered states~\cite{Cao2018,Lu2019,Sharpe2019,Yankowitz2019,Xu2018,Po2018}. It was understood early on that the high density of states alters electronic compressibility and screening of Coulomb interactions~\cite{Lian2019}. Relatively little, however, is known about how the electron-phonon (el-ph) interactions change in these systems~\cite{Wu2018,Gonzalez2019,Koshino2019b,Lian2019}. 
Here we point out that an enhancement in the electron-phonon interaction analogous to the Purcell effect in optics --- tunability of photon emission rates through changing an effective photon mode volume~\cite{Purcell1946,Scully1997} --- can be realized for electrons in superlattices (SLs) through engineering the Wannier orbital volume and shape. These effects are particularly interesting in graphene moir\'e SLs where Wannier orbitals can be tuned by the twist angle. We demonstrate that the el-ph coupling strength, enhanced by reshaping the orbitals, can be substantially larger than the values known for monolayer graphene. As an illustration of this general behavior, we present a detailed analysis of two effects that are currently under active investigation: the electron-lattice cooling and phonon-mediated linear-$T$ resistivity. We find that, in both cases, the el-ph coupling 
enhanced by Purcell-like effects translates in a dramatic enhancement of the measurable cooling and momentum relaxation rates.

Transport measurements in moir\'e graphene uncovered a surprising behavior indicating that the el-ph interaction in this system may differ substantially from that in pristine graphene. At low temperatures, moir\'e graphene hosts a rich variety of correlated states, superconducting and insulating, showing a complicated dependence on the carrier density and twist angle~\cite{Cao2018,Lu2019,Sharpe2019,Yankowitz2019}. At higher temperatures, however, this rich picture gives way to a simple ``universal'' linear-$T$ behavior: a resistivity that grows approximately linearly vs. $T$ with an abnormally steep slope~\cite{Polshyn2019}. In part, this steep $T$ dependence can be understood from an enhancement in the el-ph scattering rates due to a high density of electronic states in flat bands. However, there remains a significant departure of the measured $T$ dependence from that expected from microscopic models if the el-ph interaction strength in moir\'e graphene is taken to be the same as in pristine graphene. Strikingly, the observed $T$ dependence is almost an order of magnitude steeper than that predicted by simple estimates~\cite{Ochoa2019,Wu2019,Yudhistira2019}. This observation has led to a conjecture that the el-ph coupling is enhanced due to some not-yet-known effects~\cite{Polshyn2019}. Our work provides a simple explanation of these observations in terms of a Purcell enhancement of el-ph interaction due to the reshaping of Wannier orbitals. We also predict a similar enhancement in the electron-lattice cooling rates. The latter, if measured experimentally, will provide an independent confirmation of the Purcell enhancement mechanism.

\section{Purcell effect and Umklapp scattering}
\begin{figure}[t]
  \includegraphics[width=\columnwidth]{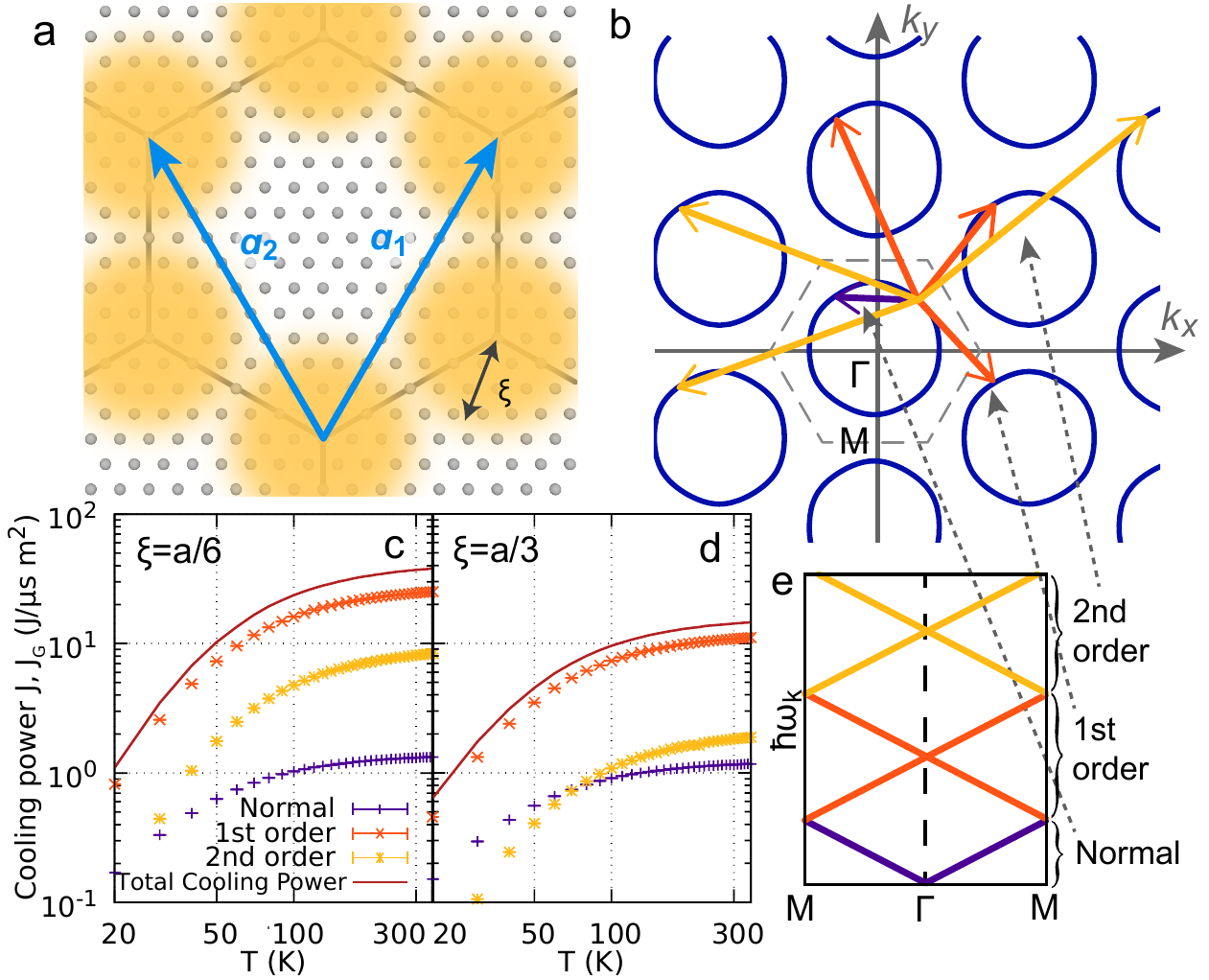}
  \caption{{\bf Electron-phonon scattering and cooling power in a superlattice.} {\bf a} The schematic figure of the model we consider. The gray balls are the atoms and the yellow cloud is the Wannier orbital of electrons. Electrons hop between the Wannier orbitals. Black lines show the unit cell of the superlattice (SL) and the two arrows are the primitive vectors of the SL. The Brillouin zone and electron-phonon scattering processes of the model are shown in {\bf b}. The dashed line in the center is the first Brillouin zone of SL and the blue circle inside is the Fermi surface. Besides the normal scattering (purple arrow), the Umklapp processes (yellow and orange arrows, see panel {\bf e}.) contribute to the scattering. The temperature dependence of cooling power (Eq.~\eqref{eq:JG}) is in {\bf c.} and {\bf d.} The two figures are for different radii of Wannier orbitals $\xi=a/6$ and $\xi=a/3$, respectively. The color of the dots corresponds to the contribution from different scattering processes in {\bf b}. The results are for $t=4.67$ meV, $a=11.7$ nm, $s=1.5\times10^4$ ms$^{-1}$, $g^2=9.37\,{\rm{eV}\AA^{2}}$, and $n=0.25\times10^{12}$ cm$^{-2}$ (see Eqs.~\eqref{eq:He}, \eqref{eq:Hp}, and \eqref{eq:g_q}). Note that the bandwidth is $W=3t=14.01$~meV. {\bf e.} is a schematic picture of the dispersion in the SL Brillouin zone. The acoustic band is folded in the 1st Brillouin zone of SL. We define the normal process as the scattering that involves the phonons in the lowest-band phonons (purple line) and the rest as the $n$th-order Umklapp process.}
  \label{fig:model}
\end{figure}

The significance of compact Wannier orbitals can be understood from a comparison to photon emission in atomic physics, which is directly proportional to the local density of optical states~\cite{Scully1997}. Replacing  electron orbitals in atoms with Wannier orbitals---and photons with  phonons---we expect the phonon emission rate to be modulated by a Wannier orbital formfactor and enhanced when the orbitals are  localized. As an illustration, we consider orbitals that are well-localized on the SL scale, such that the orbital radius $\xi$ is much smaller than the SL period $a$: $\xi \ll a$ (see schematic in Fig.~\ref{fig:model}a). Since Wannier orbitals define a formfactor that modulates the el-ph transition matrix element~\cite{Ziman1960,Koshino2019b}, the typical momentum transfer in the el-ph scattering processes will be of the order $q\sim 1/\xi$, i.e. much greater than the SL reciprocal vector. As a result, the el-ph scattering can ``outcouple" electrons in the flat bands to phonons with momenta far outside the SL Brillouin zone. Scattering by such large-momentum phonons cannot occur through the usual el-ph transitions in which phonon momentum transfer is $q\lesssim 2k_F$. The large-$q$ phonon emission enabled by compact Wannier orbitals is expected to produce a Purcell-like enhancement of the scattering rate. Indeed, theories of flat bands in moir\'e graphene predict Wannier orbitals that are well-localized on the SL  scale~\cite{Laissardiere2010,Kang2018,Koshino2018,Carr2019}. As we will see below, such compact Wannier orbitals can produce a considerable Purcell-like enhancement of el-ph scattering for phonons of wavelength $\lambda\lesssim a$.

It is  instructive to note an analogy with the Purcell-like enhancement of plasmon emission observed in experiments on nanowire systems~\cite{Akimov2007}, in which electronic orbitals confined to quantum dots were employed to boost the plasmon emission rate. Unlike the electronic states in this experiment, that were fixed-size, the Wannier orbitals in moir\'e graphene can be reshaped by changing the twist angle. This offers an appealing possibility to make Purcell enhancement tunable. 

The Purcell enhancement can also be interpreted in a simple way in terms of electron Umklapp scattering, the process in which electron momentum changes by a SL reciprocal vector as illustrated in Fig.~\ref{fig:model}. Crucially, unlike  Umklapp processes in pristine graphene monolayer, which do not occur below room $T$, the small SL period in moir\'e graphene makes these processes relevant even at low $T$. Furthermore, since the flat bands result from strong Bragg scattering of electronic waves by the SL potential, the amplitudes of electronic Umklapp scattering in a SL are expected to be relatively strong.  This mechanism will activate scattering processes that are prohibited in pristine graphene; the amplitudes of these processes are given by the Fourier harmonics of the Wannier orbital, in agreement with the discussion above.

There are many different Umklapp processes arising in this manner, some of which are illustrated in Fig.~\ref{fig:model}b. In Fig.~\ref{fig:model}e, the acoustic-phonon band is shown folded in the SL Brillouin zone; different parts of this band are accessed through different Umklapp pathways as indicated by dotted lines. The purple arrow in Fig.~\ref{fig:model}b is the momentum-conserving ${\vec G}=0$ scattering by the low-energy phonons in Fig.~\ref{fig:model}e. The processes shown in orange and yellow involve higher-energy parts of the phonon band and violate momentum conservation, i.e., ${\vec G}\ne0$. To delineate the contributions of different processes we will refer to the former processes as normal scattering (shown in purple in Fig.~\ref{fig:model}e), and those by the higher parts of the phonon band as order-$n$ Umklapp scattering ($n=1,2,3,\cdots$).

We note parenthetically that in addition to the electron Umklapps one may also consider phonon Umklapps. However, since phonon Bragg scattering is considerably weaker than electron Bragg scattering~\cite{Koshino2019}, we will focus on the electronic Umklapps and ignore the phonon Umklapps. For the sake of simplicity, we consider the longitudinal acoustic phonon band only; the electron interaction with transverse acoustic phonons are subject to a similar  Purcell enhancement that will be discussed elsewhere.

\section{Model}
To explore the effect of compact Wannier orbitals on el-ph coupling, we employ a toy model in which the nearly-flat band is described as a tight-binding problem on a honeycomb lattice with the periodicity and symmetry matching those of moir\'e SLs; Wannier orbitals are modeled as gaussian functions $w({\vec r})\sim e^{-r^2/2\xi^2}$ with the radius $\xi$ adjusted to match the realistic orbital size, as pictured in Fig.~\ref{fig:model}a. The electron Hamiltonian reads
\begin{nalign}\label{eq:He}
H_{\rm e}=&\sum_{\bf k}
\left(\begin{array}{c}
\psi_{{\vec k},1}^\dagger \\ \psi_{{\vec k},2}^\dagger
\end{array}\right)^{\rm T}
\left(\begin{array}{cc}
0 & h_{\vec k} \\
h_{\vec k}^\ast & 0
\end{array}\right)
\left(\begin{array}{c}
\psi_{{\vec k},1} \\
\psi_{{\vec k},2}
\end{array}\right),\\
h_{\vec k}=&t\left(1+e^{-i\vec k\cdot\vec a_1}+e^{-i\vec k\cdot\vec a_2}\right)
\end{nalign} 
where $t$ is the nearest-neighbor hopping, ${\vec a}_{1,2}=a\left(\pm\frac12,\frac{\sqrt3}2\right)$, $|\vec a_{1,2}|=a$, are primitive vectors and ${\vec k}=(k_x,k_y)$ is electron momentum. The phonon field is the displacement of background gray atoms placed with a distance $a_{\rm cc}\ll a$. Since temperatures of interest are much lower than graphene monolayer Debye's temperature, we focus on the $a_{\rm cc}\to0$ limit; $a_{\rm cc}$ does not appear in our calculation in this limit. The phonon Hamiltonian reads
\begin{nalign}\label{eq:Hp}
H_{\rm p}=&\sum_{\vec q, \vec G}\hbar\omega_{\vec q+\vec G}\hat a_{\vec q+\vec G}^\dagger\hat a_{\vec q+\vec G},
\quad
\omega_{\vec q+\vec G}=s|{\vec q+\vec G}|
,
\end{nalign}
where $\hbar\omega_{\vec q+\vec G}$ is the energy of the phonon with momentum $\vec q + \vec G$. The sum runs over the momenta $\vec q$ in the SL Brillouin zone and the reciprocal lattice vectors ${\vec G}=n_1{\vec G}_1+n_2{\vec G}_2$, $n_{1,2}=0,\pm 1,\pm 2...$, ${\vec G}_{1,2}=\frac{2\pi}a\left(\pm1,\frac1{\sqrt3}\right)$. Recent study of phonons in moir\'e graphene finds a dispersion that, at low energies, is similar to that of the monolayer graphene~\cite{Koshino2019,Koshino2019b}. Therefore, we used the sound velocity $s=1.5\times10^4\;{\rm ms^{-1}}$, a typical number observed in graphene~\cite{Koshino2019,Cong2019}. The electrons and phonons interact through deformation coupling,
\begin{nalign}
H_{\rm ep} = -\sum_{{\vec q},{\vec k},{\vec G},n} \frac{g_{\vec q+\vec G}}{\sqrt{V}} &\sqrt{\hbar\omega_{\vec q+\vec G}} \hat\psi^\dagger_{{ \overline{\vec k+\vec q}},n}\hat\psi_{{\vec k},n}\\
& \times (\hat a_{\vec q+\vec G}+\hat a_{-{\vec q-\vec G}}^\dagger).\label{eq:Heph}
\end{nalign}
Here, $V\equiv\sqrt3a^2N/2$ is the area of the system with $N$ being the number of the SL unit cells. We will use the notation  $\overline{\vec k+\vec q}$ to denote momenta $\vec k+\vec q+\vec G$ shifted to the first SL Brillouin zone by adjusting $\vec G$. The modulation of the electron-phonon coupling due to the Wannier orbital is described by a formfactor 
\begin{equation}\label{eq:g_q}
g_{\vec q}=g\,e^{-q^2\xi^2/4},
\end{equation}
where we assumed $g^2=9.37\,{\rm eV\AA^2}$; the value of $g$ corresponds to the deformation potential $D=20\,{\rm eV}$~\cite{Chen2008,Efetov2010} and mass density $1.52\times10^{-10}\,{\rm kg\,cm^{-2}}$. 

The ${\vec G}\ne{\vec 0}$ terms in Eq.~\eqref{eq:Heph}, which embody the enhancement of the effective el-ph coupling  due to Wannier orbital localization, appear as a consequence of the electron Bragg scattering by SL potential. In the absence of the Bragg scattering, the ${\vec G}\ne{\vec 0}$ terms are prohibited by the momentum conservation. On the other hand, when electron Bragg scattering creates well-localized Wannier function, as it does in moir\'e graphene~\cite{Kang2018,Koshino2018,Carr2019}, the momentum non-conserving $\vec G\ne\vec0$ el-ph processes become allowed. These scattering processes are illustrated in Fig.~\ref{fig:model}c. The solid circles represent a multisheet Fermi surface in the SL Brillouin zone (gray hexagon), pictured for carrier density such that the Fermi level is away from the Dirac points. The purple arrow in the SL Brillouin zone (gray hexagon) is the normal process that involves phonons with $\bf q$ in the first Brillouin zone. Besides the normal process, the Umklapp processes shown by orange and yellow arrows scatter the electron to the same final state as the normal one. The Umklapp processes contribute to the physics at  temperatures $T\gtrsim T^\ast=\hbar\omega_{\vec G}/2k_B$ ($k_B$ is Boltzmann constant) at which higher-energy phonons can be thermally activated (Fig.~\ref{fig:model}c). The quantity $T^\ast$ plays a role similar to that of the Debye temperature, however its value is much smaller than the crabin lattice Debye's temperature. For moir\'e graphene, it is $T^\ast={\cal O}(10K)$, which is close to the Bloch-Gruneisen $T$ of graphene monolayer~\cite{Koshino2019}. Therefore, the Umklapp processes contribute significantly to the electron scattering at not-too-low temperatures and, as we will see, are relevant for the temperatures of interest.

\begin{figure}[tb]
  \includegraphics[width=\columnwidth]{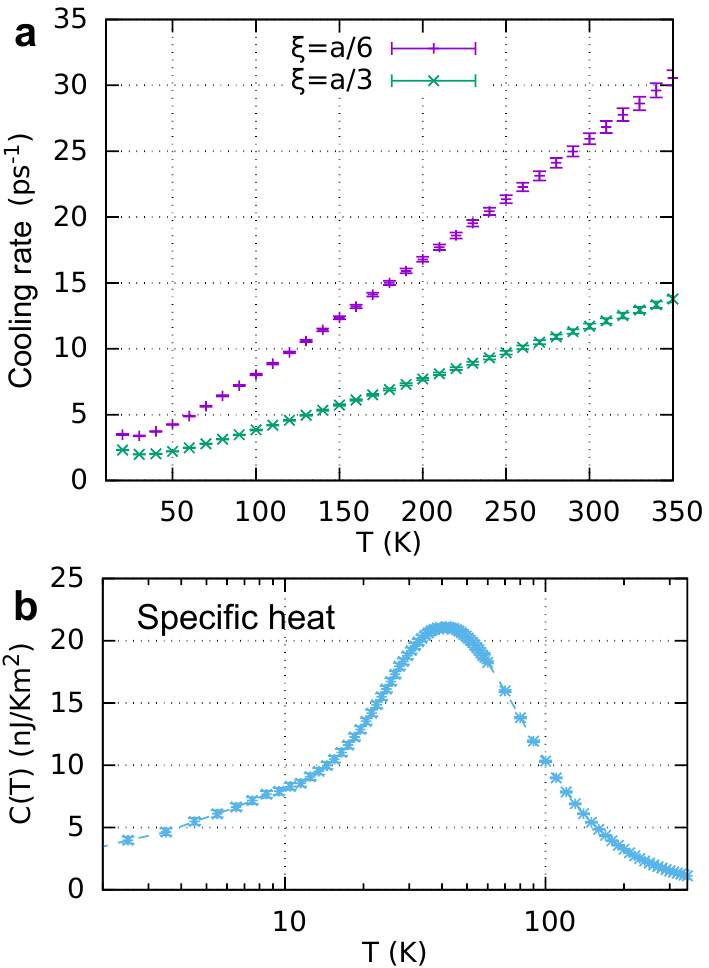}
  \caption{{\bf Electron cooling rate and heat capacity.} {\bf a.} Temperature dependence of electron cooling rate (Eq.~\eqref{eq:J'}) 
  for Wannier-function radius $\xi=a_0/6$ and $\xi=a_0/3$. {\bf b.} Temperature dependence of electron heat capacity (Eq.~\eqref{eq:C_Te}). The error bars in the two figures are the estimated errors of numerical integral. The results are for $t=4.67$ meV, $a=11.7$ nm, $s=1.5\times10^4$ ms$^{-1}$, $g^2=9.37\,{\rm eV\AA^{2}}$, $n=0.25\times10^{12}$ cm$^{-2}$, and $T_{\rm ph}=10$ K. Note that the bandwidth is $W=3t=14.01$ meV.
  }
  \label{fig:coolingrate}
\end{figure}

\section{Electron cooling}
One quantity that is highly sensitive to the nature of el-ph scattering is the electron cooling. We evaluate the cooling power using the two-temperature model~\cite{Bistritzer2009,Song2012}. This model assumes the electrons and phonons are in equilibrium with different temperatures $T_{\rm e}$ and $T_{\rm ph}$, respectively. Using Fermi's golden rule~\cite{Bistritzer2009,Song2012} and the interaction Hamiltonian $H_{\rm ep}$ from Eq.~\eqref{eq:Heph}, the cooling power per a unit area reads $J=\sum_{\vec G}J_{\vec G}$, where
\begin{align}
J_{\vec G}\equiv\sum_{n,n'}\int\frac{dk^2}{(2\pi)^2}\frac{dk'^2}{(2\pi)^2}&(\varepsilon_{n\vec k}-\varepsilon_{n\vec k'})W_{n{\vec k}',n{\vec k}}^{\vec G}f_{n\vec k}(1-f_{n\vec k'}),\label{eq:JG}
\end{align}
is the contribution from the $\vec G$ phonon band with
\begin{nalign}\label{eq:summary:Wkk}
W_{n'{\vec k}',n{\vec k}}^{\vec G}&=\delta_{n'n}2\pi|g_{\overline{\vec k'-\vec k}+{\vec G}}|^2\omega_{\overline{\vec k'-\vec k}+{\vec G}}\\
&\times\left\{N_{\overline{\vec k'-\vec k}+{\vec G}}\delta(\varepsilon_{n'\vec k'}-\varepsilon_{n\vec k}-\omega_{\overline{\vec k'-\vec k}+{\vec G}})\right.\\
&\quad+\left.(N_{\overline{\vec k'-\vec k}+{\vec G}}+1)\delta(\varepsilon_{n'\vec k'}-\varepsilon_{n\vec k}+\omega_{\overline{\vec k'-\vec k}+{\vec G}})\right\}.
\end{nalign}
Here, $N_{\vec q}=\frac1{e^{\beta_{\rm ph}sq}-1}$ is the phonon Bose-Einstein distribution and $\beta=1/T$ is the inverse temperature. The total cooling power is the sum of the contributions from different scattering processes.

Figures~\ref{fig:model}c~and~\ref{fig:model}d show the numerical result of total cooling power $J$ (solid line) and $J_{\vec G}$ for $|{\vec G}|=0$ (normal), $|{\vec G}|=\frac{4\pi}{a\sqrt3}$ (first-order Umklapp), and $|{\vec G}|=\frac{4\pi}{a}$ (second-order Umklapp). In Fig.~\ref{fig:model}c, the contribution from $J_{{\vec G}\ne{\bf0}}$ is larger than that of the normal process $J_{\vec 0}$. Indeed, $J$ is an order of magnitude larger than $J_{\vec 0}$. This trend is robust against the change of $\xi$ as seen from the $\xi=a/3$ result in Fig.~\ref{fig:model}d. The significant contribution from $J_{{\vec G}\ne{\bf0}}$ is a consequence of two factors. First, each $J_{\vec G\ne\vec0}$ gives a contribution comparable to the normal process $J_{{\vec G}={\bf0}}$ when $T_e\gtrsim T^\ast$. Second, there are six bands for each order of Umklapp processes, which enhances the contribution by almost an order of magnitude. The low frequency of the phonons and the multiplicity of $|{\vec G}|\ne0$ bands results in a large contribution.

Experimentally, the cooling power affects the electron cooling rate, i.e., the decrease of electron temperature per a unit time. The cooling rate reads
\begin{align}\label{eq:J'}
J'(T_{\rm e},T_{\rm ph})=\frac{J(T_{\rm e},T_{\rm ph})}{C(T_{\rm e})\Delta T},
\end{align}
where $\Delta T\equiv T_e-T_{ph}$, $C(T_e)$ is the heat capacity, and $T_{\rm e}$ and $T_{\rm ph}$ are respectively the temperatures of electrons and phonons. We calculate the heat capacity using
\begin{align}\label{eq:C_Te}
C(T_{\rm e})=&\sum_n\int_{1BZ}\frac{dk^2}{(2\pi)^2}\frac{4k_B\beta_{\rm e}^2\varepsilon_{n{\vec k}}^2}{\cosh^2(\beta_{\rm e}\varepsilon_{n\vec k}/2)},
\end{align}
because the temperature is comparable to the bandwidth. Figure~\ref{fig:coolingrate} shows the cooling rate [Fig.~\ref{fig:coolingrate}a] and the electron heat capacity [Fig.~\ref{fig:coolingrate}b] between 1 K $\le T_{\rm e}\le$ 350 K. The heat capacity $C(T_{\rm e})$ shows a Schottky peak at $T_{\rm e}\sim20$ K, which manifests a crossover from the low-temperature ($k_BT_{\rm e}\ll t$) to high-temperature ($k_BT_{\rm e}\gg t$) regime. Interestingly, the cooling rate $J'(T_{\rm e},T_{\rm ph})\sim J/C$, pictured in Fig.~\ref{fig:coolingrate}a, increases monotonically and approximately linearly with $T_{\rm e}$ showing no peak, a behavior that reflect the decrease of $C(T_{\rm e})$ at the high temperature.



The $T_{\rm e}$-linear dependence of $J'(T_{\rm e},T_{\rm ph})$ in the high-$T_e$ is a manifestation of the asymptotic behavior in the high-$T_{\rm e}$ limit. In the limit $k_{\rm B}T_{\rm e}\gg k_{\rm B}T_{\rm ph}, \hbar\omega_{\vec G}, t$, the cooling power reads
\begin{align}
J(T_{\rm e},T_{\rm ph})\sim\frac{2\pi}\hbar\sum_{n}\int_{1BZ}\frac{dk^2}{(2\pi)^2}\eta_{n{\vec k},{\vec G}},
\end{align}
where
\begin{nalign}
\eta_{n{\vec k}}\equiv&\sum_{\vec G}\int_{1BZ}\frac{dk'{}^2}{(2\pi)^2}(\hbar\omega_{\overline{{\vec k}'-{\vec k}}+{\vec G}})^2|g_{n,\overline{{\vec k}'-{\vec k}}+{\vec G}}|^2\\
&\hspace{2cm}\times\delta\left(\varepsilon_{n{\vec k}'}-\varepsilon_{n{\vec k}}-\hbar\omega_{\overline{{\vec k}'-{\vec k}}+{\vec G}}\right),
\end{nalign}
and
\begin{align}
C(T_e)=&\frac1{k_{\rm B}T_{\rm e}^2}\sum_n\int_{1BZ}\frac{dk^2}{(2\pi)^2}\varepsilon_{n{\vec k}}^2.
\end{align}
The approximately linear $T$-dependence $J'(T_{\rm e},T_{\rm ph})\propto T_{\rm e}$, which is valid even in the high-$T_{\rm e}$ limit, arises because 
$J\propto{\rm const.}$ and $C\propto T_{\rm e}^{-2}$ at high $T$. 



\begin{figure}[tb]
  \centering
  \includegraphics[width=\columnwidth]{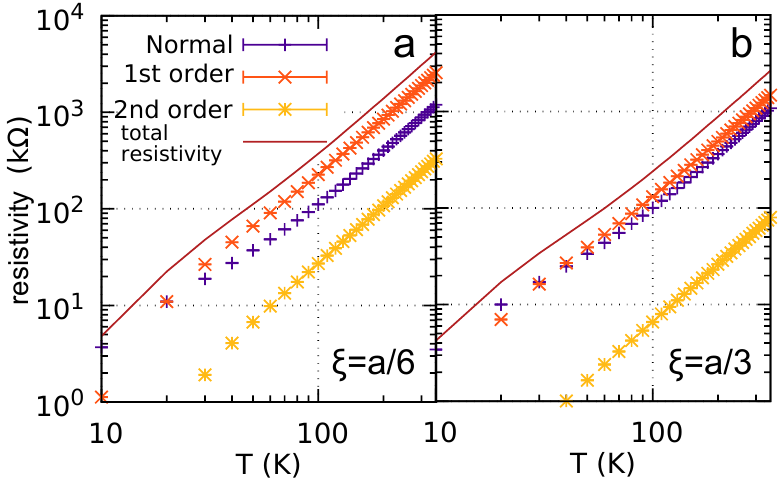}
  \caption{{\bf Temperature dependence of resistivity $\rho(T)$ and the contributions from different scattering processes.} The two figures are for different Wannier orbital radius $\xi$: {\bf a.} $\xi=a/6$ and {\bf b.} $\xi=a_0/3$. The solid line shows the resistivity and the points are the contribution from normal, first-order Umklapp, and second-order Umklapp processes (Eq. \eqref{eq:rho_G}). The results are for $t=4.67$ meV, $a=11.7$ nm, $s=1.5\times10^4$ ms$^{-1}$, $g^2=9.37\,{\rm eV\AA^{2}}$ and $n=0.25\times10^{12}$ cm$^{-2}$. Note that the bandwidth is $W=3t=14.01$ meV.}
  \label{fig:rho}
\end{figure}

\section{Resistivity}
Umklapp processes also contribute to the electrical resistivity. The total resistivity $\rho$ of a system is related to the entropy production by the electron-phonon scattering. We consider only the longitudinal current as the transverse current is prohibited by symmetry. Within the variational method assuming the modulation of electron distribution $\delta f_{n\vec k}\equiv f_{n\vec k}-f_{n\vec k}^0=\tau e\vec E\cdot\vec v_{n\vec k}\beta f^0_{n{\vec k}}(1-f^0_{n{\vec k}})$~\cite{Ziman1960}, $\rho(T)$ follows Matthiessen's rule $\rho(T)=\sum_{\vec G}\rho^{\vec G}(T)$ where
\begin{align}\label{eq:rho_G}
\rho^{\vec G}(T)=&\frac{\sum_{n,n'}\int\frac{dk^2}{(2\pi)^2}\frac{dk'^2}{(2\pi)^2}\beta P^{\vec G}_{n{\vec k},{\vec k'-\vec k};n'{\vec k}'}\left(v^x_{n{\vec k}}-v^x_{n'{\vec k}'}\right)^2}{\left[e\sum_{n}\int\frac{dk^2}{(2\pi)^2}(v_{n{\vec k}}^x)^2\beta f^0_{n{\vec k}}(1-f^0_{n{\vec k}})\right]^2},
\end{align}
are the contributions of the phonon $\vec G$ bands with
\begin{align}
P_{n{\vec k},{\vec q};n'{\vec k'}}^{\vec G}=&2\pi\delta_{n,n'}|g_{\vec q+\vec G}|^2\omega_{\vec q+\vec G} N_{\vec q+\vec G}f^0_{n{\vec k}}\left(1-f^0_{n'{\vec k'}}\right)\nonumber\\
&\qquad\times\delta(\varepsilon_{n{\vec k}}+\hbar\omega_{\vec q+\vec G}-\varepsilon_{n'{\vec k}'}).
\end{align}
Here, $\vec E$ is the external electric field, and $\tau$ is the variational parameter. We assume $T=T_{\rm e}=T_{\rm ph}$. The temperature dependence of $\rho$ and $\rho^{\bf G}$ is shown in Fig.~\ref{fig:rho}. The main contribution comes from the first-order Umklapp process at temperatures $T\gtrsim T^\ast$, which is larger than that of the normal process. 

\section{Electron-phonon coupling in moir\'e graphene}
\begin{figure}[t]
  \centering
  \includegraphics[width=\columnwidth]{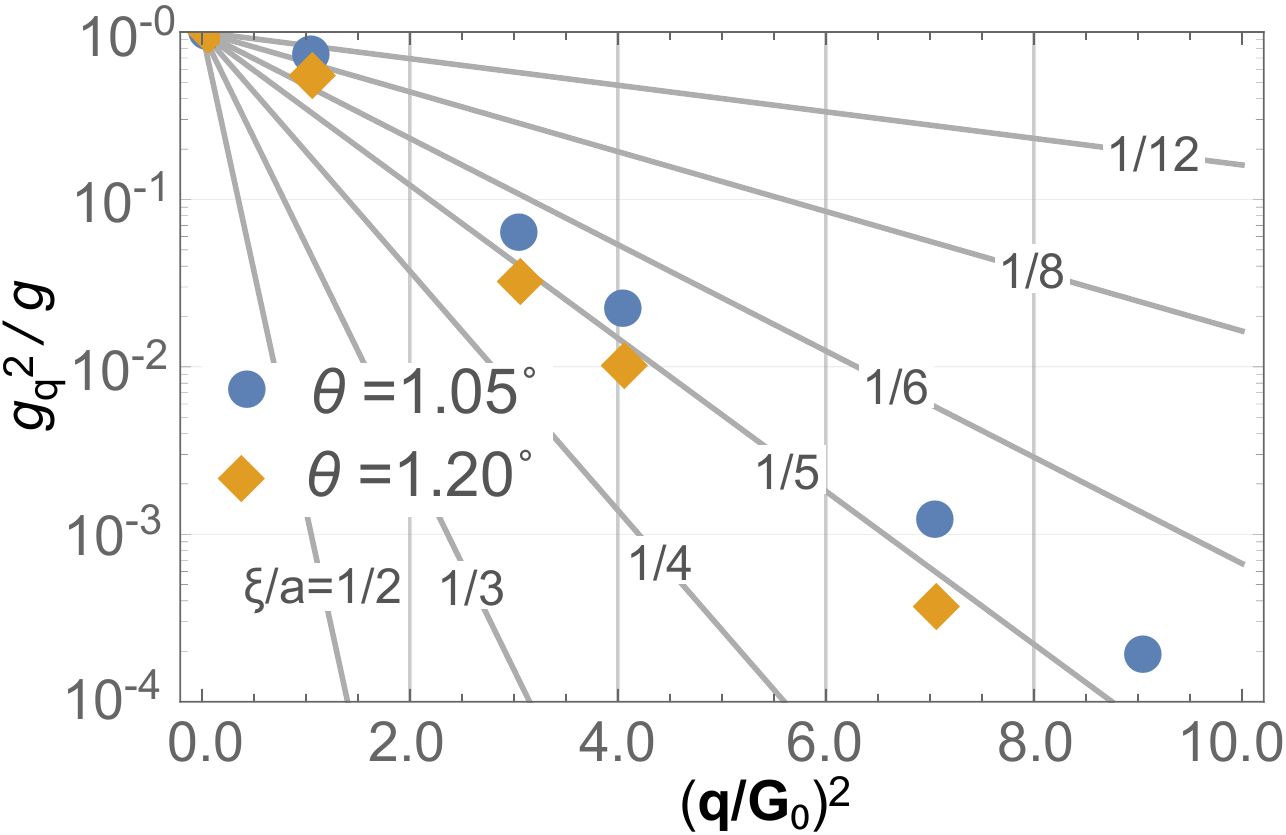}
  \caption{{\bf The electron-phonon coupling form factor $g_{\bf q}$ for moir\'e graphene (Eq.~\eqref{eq:g_q}).} The disks and diamonds show the form factor for twist angles $\theta=1.05^\circ$ and $1.20^\circ$, respectively. They are calculated from the Wannier functions of the continuum model (See Supplementary Information). The gray lines are the gaussian form factor $g_{\bf q}\propto\exp(-\xi^2q^2/4)$ with different Wannier function radius $\xi$.}
  \label{fig:gq}
\end{figure}

To validate the conclusions drawn from the toy model, we made a detailed comparison to the microscopic continuum model of moir\'e graphene~\cite{Bistritzer2011} (See Supplementary Information for technical details). The results for different Umklapp processes, shown in Fig.~\ref{fig:gq}, are in good agreement with the toy model that links the Umklapp scattering rates to the Wannier function formfactor $g_q$. As Fig.~\ref{fig:gq} illustrates, the best fit to the microscopic calculation is indeed provided by tightly localized Wannier orbitals as in Eq.~\eqref{eq:g_q} with $\xi/a\sim1/8-1/5$, supporting the Purcell-like enhancement of el-ph scattering rates due to wannier orbitals localization. Furthermore, this best-fit value appears to be independent of the twist angle, adding an air of confidence to the toy model approach.

\section*{Discussion}
We see that reshaping of Wannier orbitals impacts the observables such as the cooling power and resistivity. The enhancement in the el-ph scattering rates due to Wannier orbital compression originates from the same physics as the Purcell effect in quantum optics. The enhancement has a nontrivial dependence on phonon momentum, being particularly important for large-$q$ phonons. As the analysis above indicates, these phonons are responsible for a substantial enhancement of the cooling power and the linear-$T$ resistivity. The exact value for such an enhancement will depend on system details on the SL scale;  for realistic parameter values it can be as high as tenfold, or more. 

A key parameter that controls the enhancement is the Wannier orbital radius $\xi$ which suppresses the electron-phonon scattering for $q\xi\gtrsim1$ (Eqs.~\ref{eq:Heph},~\ref{eq:g_q}).  Shrinking of Wannier orbitals, described by a reduction in $\xi$, leads to a stronger effective el-ph coupling arising because the large-$q$ phonons with $|q|\gtrsim\pi/a$ can now be activated. Indeed,   the el-ph formfactor $g_q\sim e^{-q^2\xi^2}$, Eq.~\ref{eq:g_q}, is small for large-$q$ phonons $|q|\gtrsim\pi/a$ when the Wannier orbitals extend throughout the SL unit cell, i.e., $\xi \sim a$, but increases as the Wannier orbitals shrink, $\xi\ll a$. This is exactly what happens in moir\'e graphene due to its localized Wannier orbitals~\cite{Laissardiere2010,Kang2018,Koshino2018,Carr2019}.

The tunability of the el-ph interactions of moir\'e materials through reshaping the Wannier orbitals has many interesting implications. The dramatic enhancement of the cooling and momentum relaxation rates, the two effects that  are straightforward to measure experimentally, will offer an independent confirmation of the Purcell enhancement mechanism. The connection between enhanced interactions and reshaped Wannier orbitals will provide unique means to control interactions in moir\'e bands. As a new direction in the study of moir\'e materials, it will facilitate the microscopic understanding of superconductivity and other strongly-correlated phenomena.


We thank Andrey Shytov for useful discussions. HI was supported by JSPS KAKENHI (Grant Numbers JP18H03676, JP18H04222, and JP19K14649) and UTokyo Global Activity Support Program for Young Researchers. LL acknowledges support from the Science and Technology Center for Integrated Quantum Materials, NSF Grant No. DMR-1231319; and Army Research Office Grant W911NF-18-1-0116. FG was supported by funding from the European Commission, under the Graphene Flagship, Core 3, grant number 881603, and by the grants NMAT2D (Comunidad de Madrid, Spain),  SprQuMat and SEV-2016-0686, (Ministerio de Ciencia e Innovaci\'on, Spain).



\begin{thebibliography}{99}
\bibitem{LopesDosSantos2007}     Lopes dos Santos, J. M. B., Peres, N. M. R. \& Castro Neto, A. H. Graphene bilayer with a twist: electronic structure. Phys. Rev. Lett. {\bf99}, 256802 (2007).
\bibitem{Laissardiere2010}       Trambly de Laissardi\'ere, G., Mayou, D. \& Magaud, L. Localization of Dirac electrons in rotated graphene bilayers. Nano Lett. {\bf10}, 804 (2010).
\bibitem{Mele2010}               Mele, E. J. Commensuration and interlayer coherence in twisted bilayer graphene. Phys. Rev. B {\bf81}, 161405(R) (2010).
\bibitem{Shallcross2010}         Shallcross, S., Sharma, S., Kandelaki, E. \& Pankratov O, A. Electronic structure of turbostratic graphene. Phys. Rev. B {\bf81}, 165105 (2010).
\bibitem{Morell2010}             Su\'arez Morell, E., Correa, J. D., Vargas, P., Pacheco, M. \& Barticevic, Z. Flat bands in slightly twisted bilayer graphene: Tight-binding calculations. Phys. Rev. B {\bf82}, 121407 (2010).
\bibitem{Bistritzer2011}         Bistritzer, R. \& MacDonald, A. H. Moir\'e bands in twisted double-layer graphene. Proc. Nat. Acad. Soc. {\bf108}, 12233 (2011).
\bibitem{Cao2018}               Cao, Y., Fatemi, V., Fang, S., Watanabe, K., Taniguchi, T., Kaxiras, E. \& Jarillo-Herrero, P. Correlated insulator behaviour at half-filling in magic-angle graphene superlattices. Nature {\bf556}, 43-50 (2018). 
\bibitem{Lu2019}                 Lu, X., Stepanov, P., Yang, W., Xie, M., Aamir, M. A., Das, I., Urgell, C., Watanabe, K., Taniguchi, T., Zhang, G., Bachtold, A., MacDonald, A. H. \& Efetov, D. K. Superconductors, orbital magnets and correlated states in magic-angle bilayer graphene. Nature {\bf574}, 653-657 (2019).
\bibitem{Sharpe2019}             Sharpe, A. L., Fox, E. J., Barnard, A. W., Finney, J., Watanabe, K., Taniguchi, T., Kastner, M. A. \& Goldhaber-Gordon, D. Emergent ferromagnetism near three-quarters filling in twisted bilayer graphene. Science {\bf365}, 605-608 (2019).
\bibitem{Yankowitz2019}          Yankowitz, M., Chen, S., Polshyn, H., Zhang, Y., Watanabe, K., Taniguchi, T., Graf, D., Young, A. F. \& Dean, C. R. Tuning superconductivity in twisted bilayer graphene.Science {\bf363}, eaav1910 (2019).
\bibitem{Xu2018}                 Xu, C. \& Balents, L. Topological Superconductivity in Twisted Multilayer Graphene. Phys. Rev. Lett. {\bf121}, 087001 (2018).
\bibitem{Po2018}                 Po, H. C., Zou, L., Vishwanath, A. \& Senthil, T. Origin of Mott Insulating Behavior and Superconductivity in Twisted Bilayer Graphene. Phys. Rev. X {\bf8}, 031089 (2018).
\bibitem{Lian2019}               Lian, B., Wang, Z. \& Bernevig, B. A. Twisted bilayer graphene: A phonon-driven superconductor. Phys. Rev. Lett. {\bf122}, 257002 (2019).
\bibitem{Wu2018}                 Wu, F., MacDonald, A. H. \& Martin, I. Theory of phonon-mediated superconductivity in twisted bilayer graphene. Phys. Rev. Lett. {\bf121}, 257001 (2018).
\bibitem{Gonzalez2019}           Gonz\'alez, J. \& Stauber, T. Kohn-Luttinger superconductivity in twisted bilayer graphene. Phys. Rev. Lett. {\bf122}, 026801 (2019).
\bibitem{Koshino2019b}           Koshino, M. \& Nam, N. N. T. Continuum model for relaxed twisted bilayer graphenes and moir\'e electron-phonon interaction. Preprint (arXiv:1909.10786) (2019).
\bibitem{Purcell1946}            Purcell, E. M. Spontaneous emission probabilities at radio frequencies. Phys. Rev. {\bf69}, 681 (1946).
\bibitem{Scully1997} M. O. Scully and M. S. Zubairy, Quantum Optics
(Cambridge University Press, Cambridge, 1997).
\bibitem{Polshyn2019}        Polshyn, H., Yankowitz, M., Chen, S., Zhang, Y., Watanabe, K., Taniguchi, T., Dean, C. R. \& Young, A. F. Large linear-in-temperature resistivity in twisted bilayer graphene. Nat. Phys. {\bf15}, 1011-1016 (2019).
\bibitem{Ochoa2019}          Ochoa, H. moir\'e-pattern fluctuations and electron-phason coupling in twisted bilayer graphene. Phys. Rev. B {\bf100}, 155426 (2019).
\bibitem{Wu2019}             Wu, F., Hwang, E. \& Das Sarma, S. Phonon-induced giant linear-in-$T$ resistivity in magic angle twisted bilayer graphene: Ordinary strangeness and exotic superconductivity. Phys. Rev. B {\bf99}, 165112 (2019).
\bibitem{Yudhistira2019}     Yudhistira, I., Chakraborty, N., Sharma, G., Ho, D. Y. H., Laksono, E., Sushkov, O. P., Vignale, G. \& Adam, S. Gauge-phonon dominated resistivity in twisted bilayer graphene near magic angle. Phys. Rev. B {\bf99}, 140302 (2019).
\bibitem{Ziman1960}          Ziman, J. M. {\it Electrons and Phonons}. (Clarendon Press, 1960).
\bibitem{Kang2018}           Kang, J. \& Vafek, O. Symmetry, Maximally Localized Wannier States, and a Low-Energy Model for Twisted Bilayer Graphene Narrow Bands. Phys. Rev. X {\bf8}, 031088 (2018).
\bibitem{Koshino2018}        Koshino, M., Yuan, N. F. Q., Koretsune, T., Ochi, M., Kuroki, K. \& Fu, L. Maximally Localized Wannier Orbitals and the Extended Hubbard Model for Twisted Bilayer Graphene. Phys. Rev. X {\bf8}, 031087 (2018).
\bibitem{Carr2019}           Carr, S., Fang, S., Zhu, Z. \& Kaxiras, E. Exact continuum model for low-energy electronic states of twisted bilayer graphene. Phys. Rev. Res. {\bf1}, 013001 (2019).
\bibitem{Akimov2007}         Akimov, A. V., Mukherjee, A., Yu, C. L., Chang, D. E., Zibrov, A. S., Hemmer, P. R., Park, H. \& Lukin, M. D. Generation of single optical plasmons in metallic nanowires coupled to quantum dots. Nature {\bf450}, 402 (2007).
\bibitem{Koshino2019}        Koshino, M. \& Son, Y.-W. Moir\'e phonons in twisted bilayer grahene. Phys. Rev. B {\bf100}, 075416 (2019).
\bibitem{Cong2019}           Cong, X., Li, Q.-Q., Zhang, X., Lin, M.-L., Wu, J.-B., Liu, X.-L., Venezuela, P., Tan, P.-H. Probing the acoustic phonon dispersion and sound velocity of graphene by Raman spectroscopy. Carbon {\bf149}, 19-24 (2019).
\bibitem{Chen2008}           Chen, J.-H., Jang, C., Xiao, S., Ishigami, M. \& Fuhrer, M. S. Intrinsic and extrinsic performance limits of graphene devices on SiO$_2$. Nat. Nano. {\bf3}, 206-209 (2008).
\bibitem{Efetov2010}         Efetov, D. K. \& Kim, P. Controlling Electron-Phonon Interactions in Graphene at Ultrahigh Carrier Densities. Phys. Rev. Lett. {\bf105}, 256805 (2010).
\bibitem{Bistritzer2009}     Bistritzer, R. \& MacDonald, A. H. Electronic cooling in graphene. Phys. Rev. Lett. {\bf102}, 206410 (2009).
\bibitem{Song2012}           Song, J. C. W., Reizer, M. Y. \& Levitov, L. S. Disorder-assisted electron-phonon scattering and cooling pathways in graphene. Phys. Rev. Lett. {\bf109}, 106602 (2012).
\end{thebibliography}
\end{document}


\title{Supplemental information for\\ Tunable electron-phonon interactions in long-period superlattices}

\maketitle


\maketitle

\section{Electron-phonon coupling for continuum model}

\begin{figure}[b]
\begin{tabular}{cc}
\includegraphics[width=3.in]{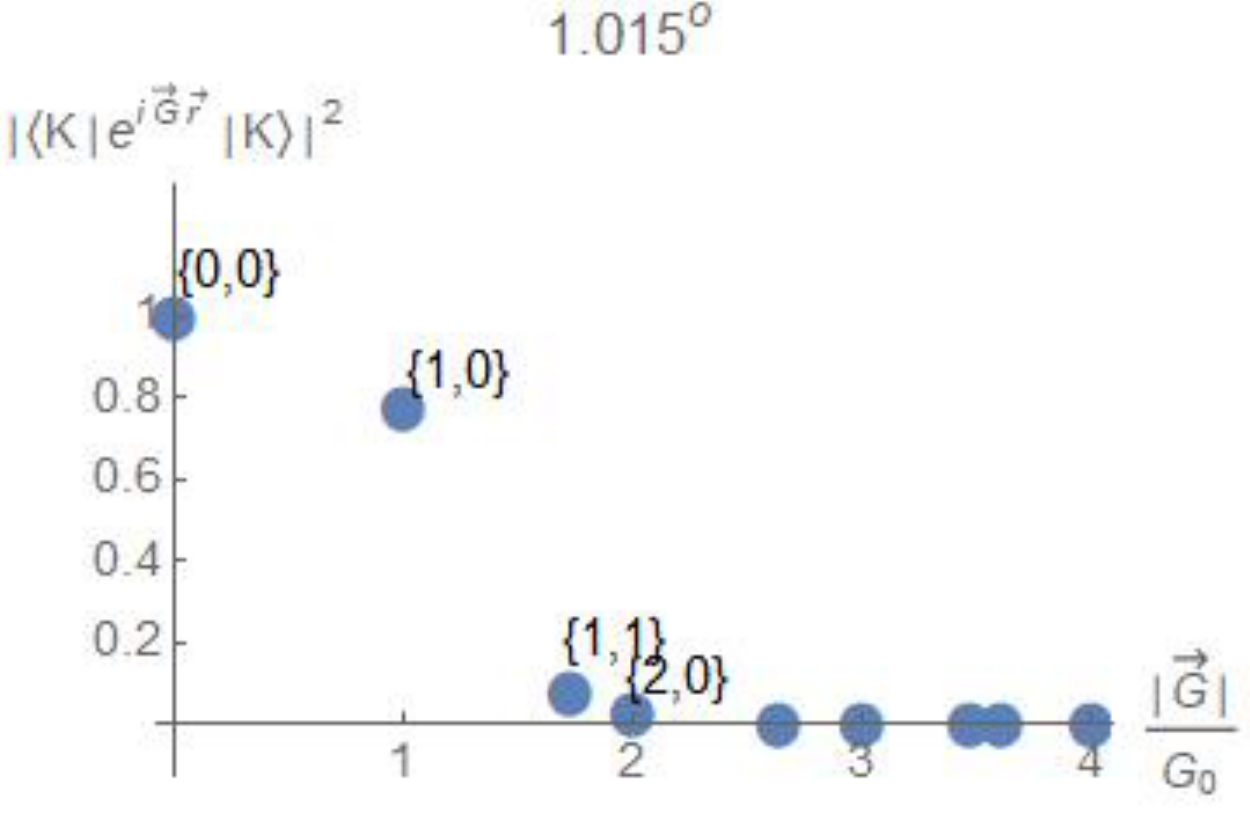} &
\includegraphics[width=3.in]{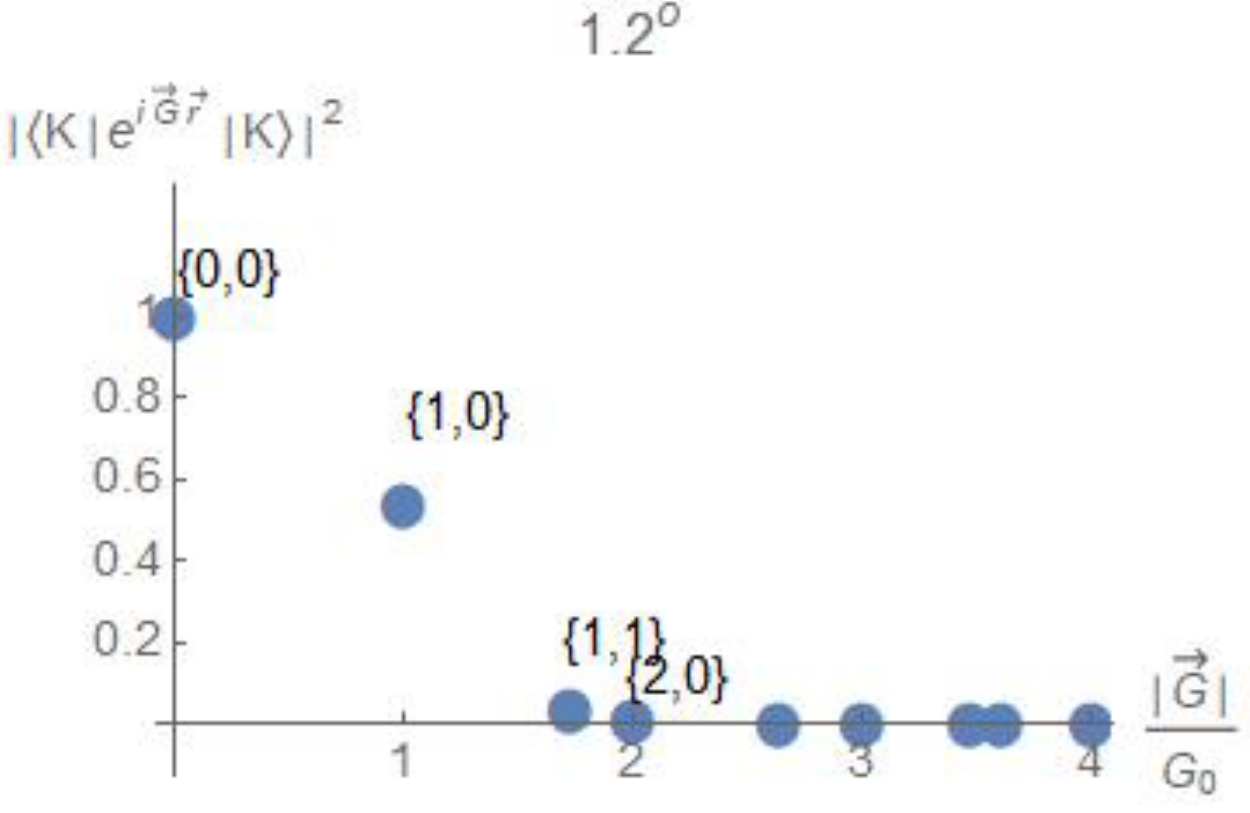}
\end{tabular}
\caption{Matrix elements squared, see Eq.~\eqref{mat2}, for Umklapp processes involving different reciprocal lattice vectors. The reciprocal lattice vectors are labeled by two integers, $\{ m , n \}$ such that $\vec{G} = m \vec{G}_1 + n \vec{G}_2$, where $\vec{G}_1$ and$\vec{G}_2$ are the unit vectors of the reciprocal lattice. It is assumed that the atomic displacements induced by the phonon are equal in both layers.}
\label{fig:mat}
\end{figure}

The electron-phonon coupling in graphene includes a scalar contribution, described by the deformation potential, and a gauge contribution, due to the modulation of the nearest neighbor electron hoppings induced by the phonons. We consider first the scalar coupling, discussed in the main text. In the twisted bilayer, the coupling between states in the narrow bands with momenta $\vec{k}$ and $\vec{k} + \vec{q}$ through a longitudinal acoustic phonon of momentum $\vec{q} + \vec{G}$, where $\vec{G}$ is a reciprocal lattice vector,  can be expressed as 
\begin{align}
g_{\vec{k}, \vec{k} + \vec{q}} &= D \left\langle \psi_{\vec{k} + \vec{q}} \left| {\cal I}_\sigma {\cal I}_\tau e^{i ( \vec{q} + \vec{G} ) \vec{r}}  \right| \psi_{\vec{k}} \right\rangle
\label{mat}
\end{align}
where the wavefunction $\left|\psi_\vec{k} \right>$ is a four component spinor, whose components are labeled by the sublattice, $\sigma$, and layer, $\tau$. Note that the full electron-phonon coupling includes a factor proportional to $| \vec{q} + \vec{G} |$. The matrices ${\cal I}_\sigma$ and $ {\cal I}_\tau$ are $2 \times 2$ unit matrices in sublattice and layer indices. The use of the matrix $ {\cal I}_\tau$ implies that we consider phonons with displacements which are equal in both layers.  In order to estimate the relative strength of Umklapp scattering processes, we analyze the square of the matrix elements, which is the quantity which appears in the transport coefficients. We calculate the dimensionless quantity.
\begin{align}
M_{\vec{K} , \vec{G} } &=  \left| \left\langle \psi_{\vec{K}} \left| {\cal I}_\sigma {\cal I}_\tau e^{i  \vec{G}  \vec{r}}  \right| \psi_{\vec{K}} \right\rangle \right|^2
\label{mat2}
\end{align}
where, in addition, we have used the states at the corners of the Brillouin Zone, $\vec{K}$, as representatives of the $\vec{G}$ dependence of the matrix elements.  By construction, $M_{K , \vec{G} = 0} = 1$. For $\vec{G} \ne 0$, we average over equivalent vectors. The calculations are using the parameters in\cite{Koshino2018s} and an angle $\theta =1.015^\circ$.Results are shown in Fig.~\ref{fig:mat}.

\begin{figure}[h]
\begin{tabular}{cc}
\includegraphics[width=3.in]{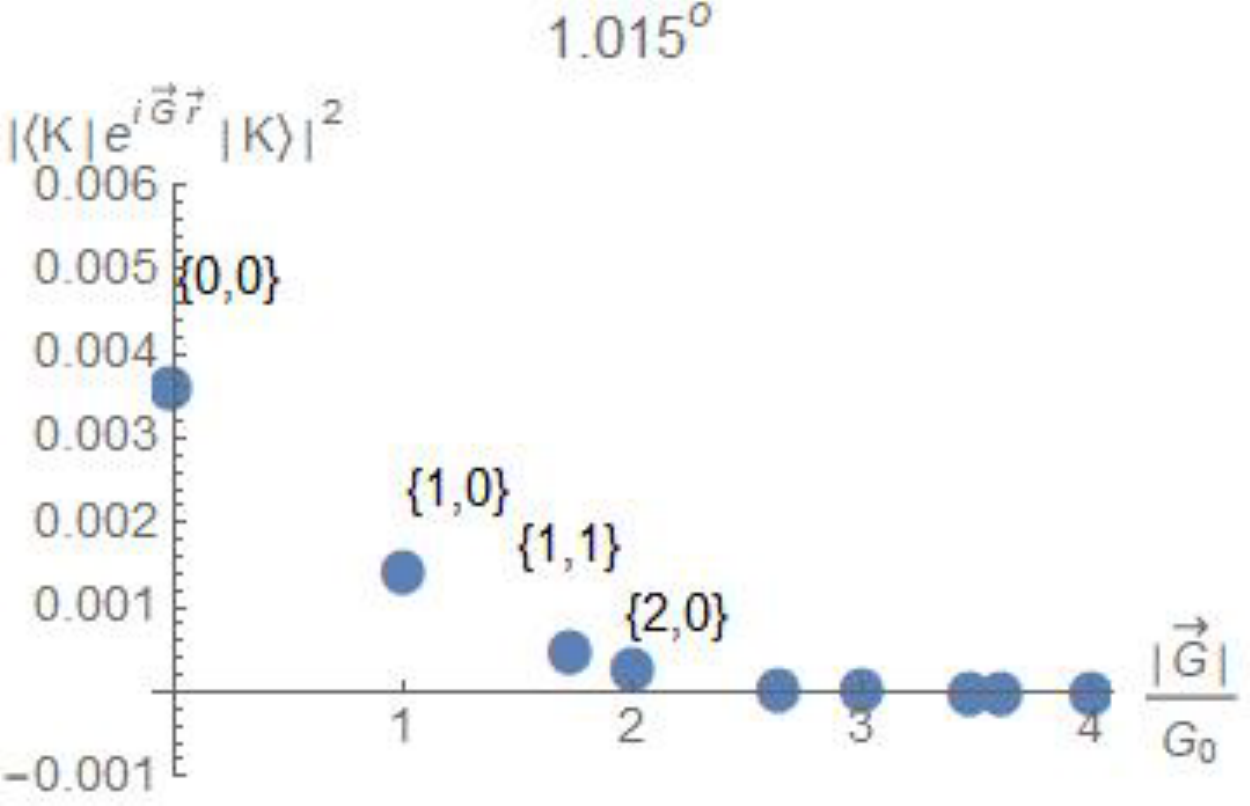} &
\includegraphics[width=3.in]{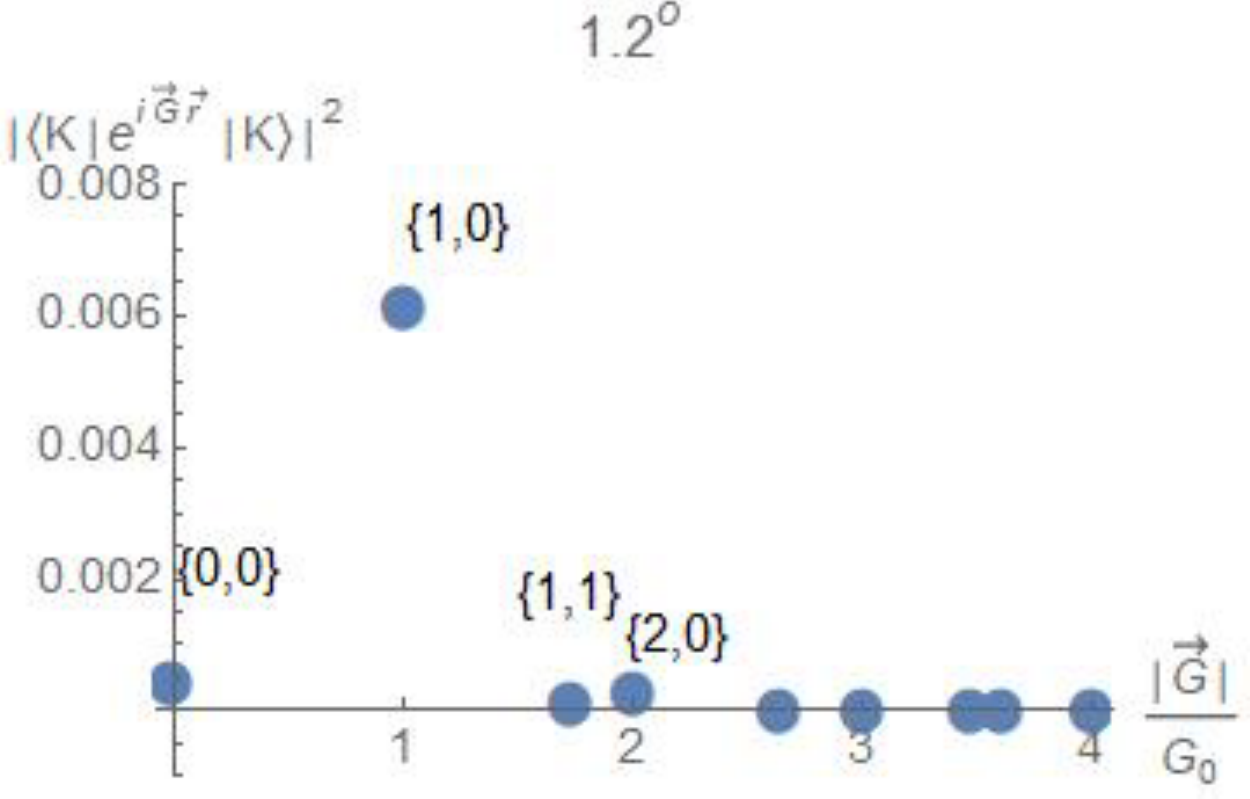}
\end{tabular}
\caption{As in Fig.~\ref{fig:mat}, but assuming that the atomic displacements have opposite signs in the two layers.}
\label{fig:mat_2}
\end{figure}

The calculation can be extended to phonons whose displacements are opposite in the two layers. We have checked that the corresponding couplings squared are, at least, one order of magnitude smaller than the ones discussed previously, see Fig.~\ref{fig:mat_2}.

\begin{figure}[h]
\begin{tabular}{cc}
\includegraphics[width=3.in]{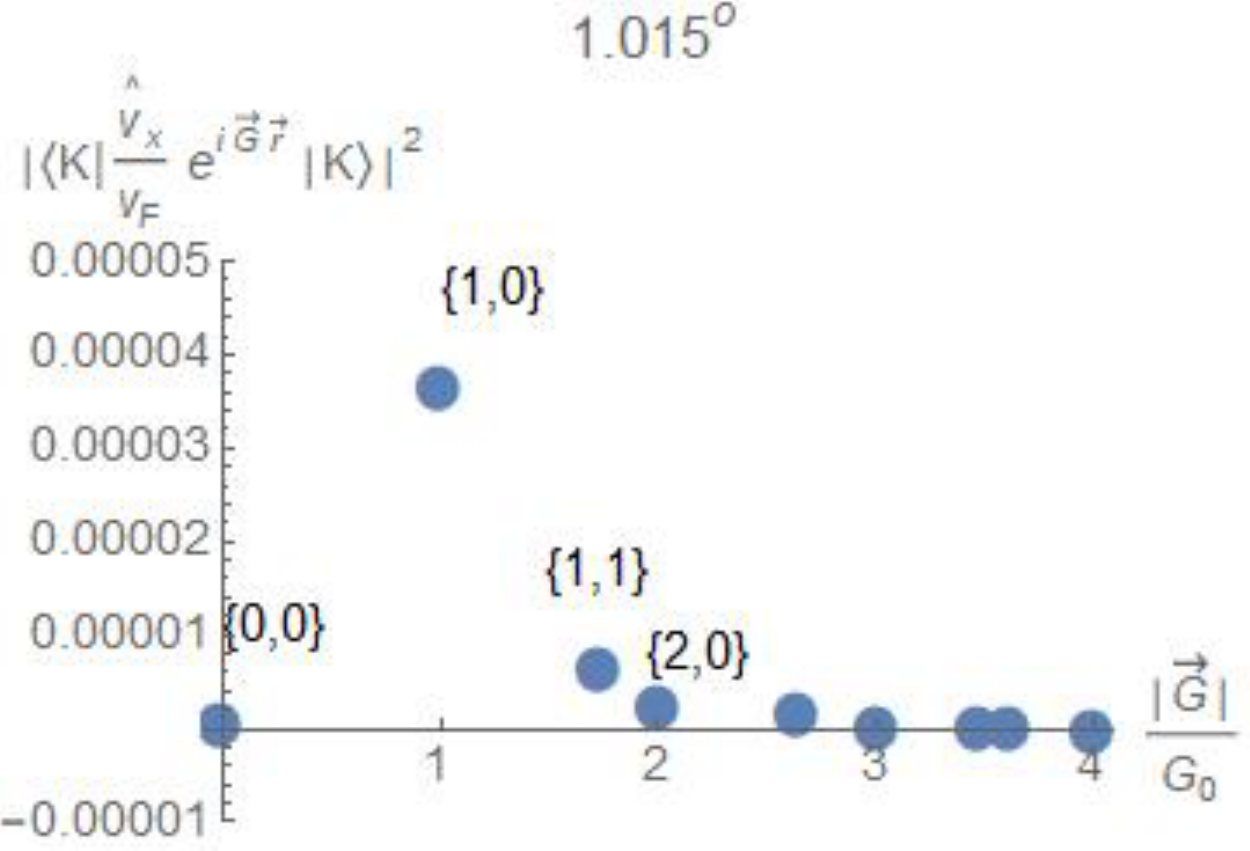} &
\includegraphics[width=3.in]{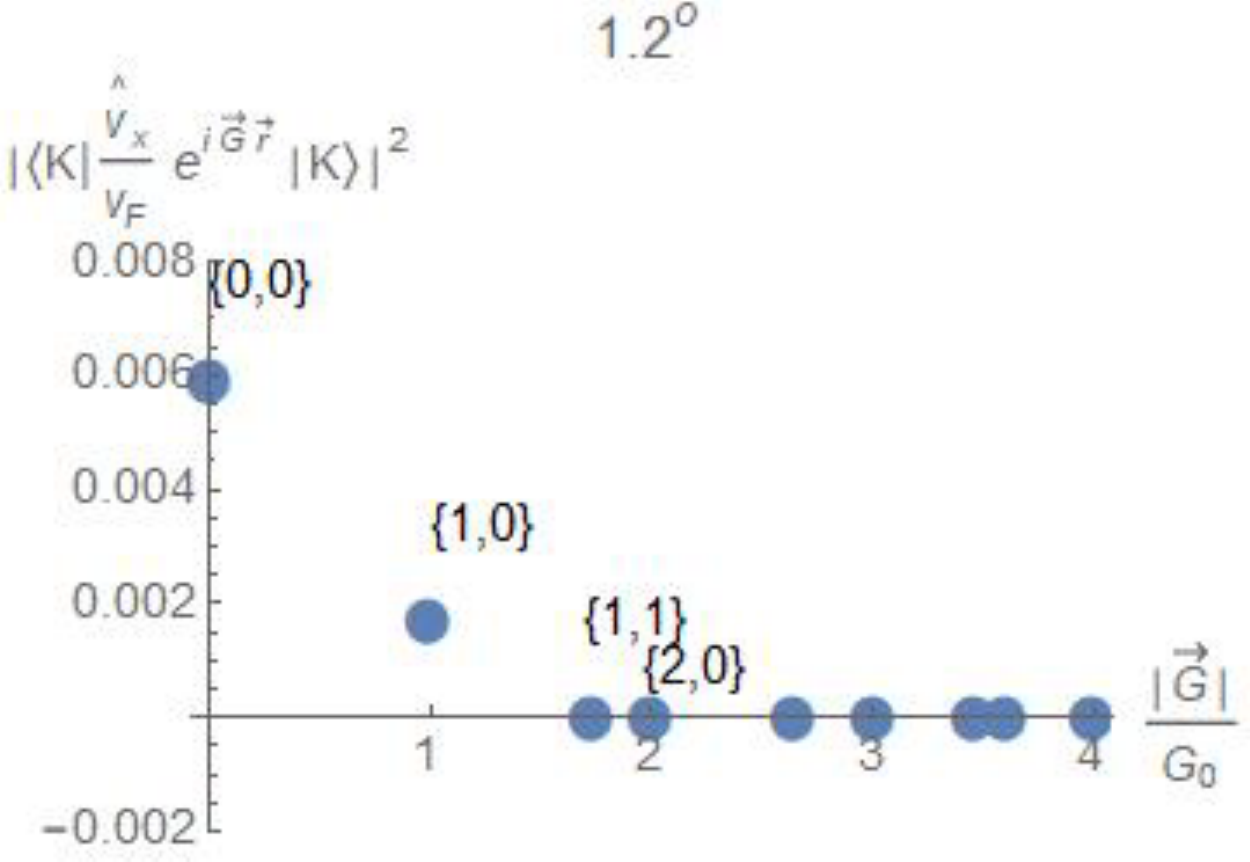}
\end{tabular}
\caption{As in Fig.~\ref{fig:mat}, but replacing the operator ${\cal I}_\sigma$ by $\sigma_x$ in Eq.~\eqref{mat} (gauge coupling, and assuming that the atomic displacements have the same sign in the two layers.}
\label{fig:mat_3}
\end{figure}

\begin{figure}[h]
\begin{tabular}{cc}
\includegraphics[width=3.in]{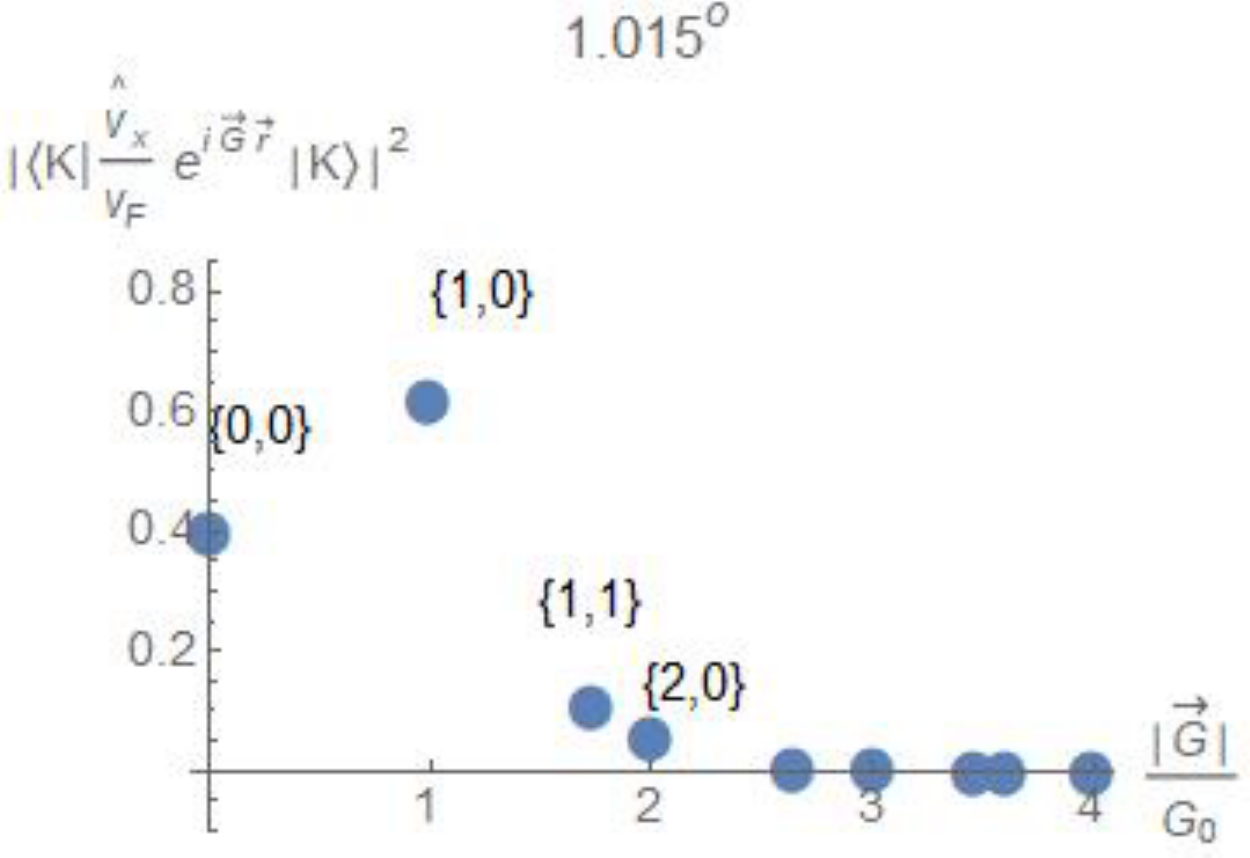} &
\includegraphics[width=3.in]{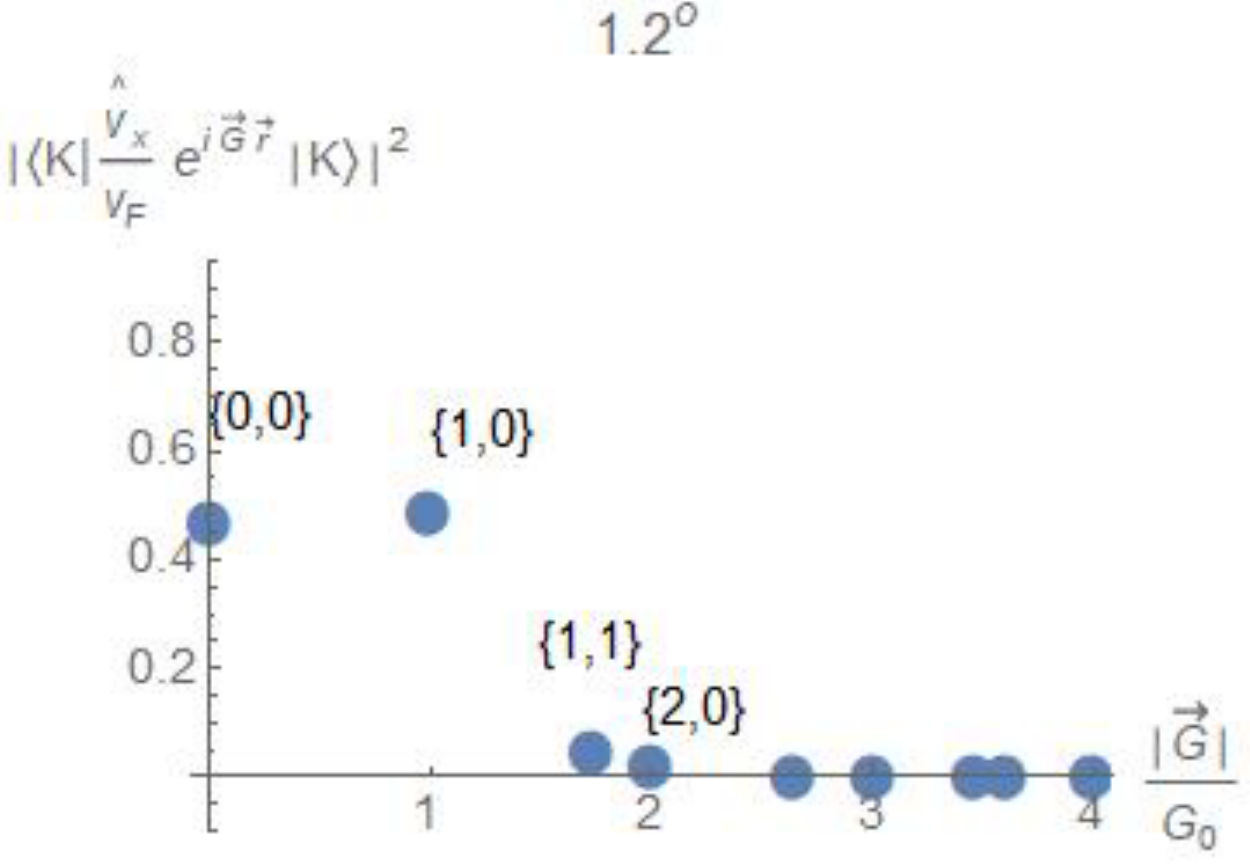}
\end{tabular}
\caption{As in Fig.~\ref{fig:mat}, but replacing the operator ${\cal I}_\sigma$ by $\sigma_x$ in Eq.~\eqref{mat} (gauge coupling, and assuming that the atomic displacements have opposite sign in the two layers.}
\label{fig:mat_4}
\end{figure}

For completeness, we have also considered the coupling to transverse acoustic phonons, where the coupling is proportional to the in-layer velocity operators, $v_x , v_y \propto \sigma_x , \sigma_y$. For a twisted graphene bilayer, the electronic states in the narrow bands are a superposition of Bloch states in both layers with almost opposite velocities\cite{Bistritzer2011s}. Hence, the coupling to transverse phonons with equal displacements in the two layers is suppressed by a factor proportional to the reduction in the Fermi velocity. This suppression does not take place if the displacements have opposite signs in the two layers. These antisymmetric phonons can contribute to the resistivity and the electron cooling. As in the case of longitudinal phonons, a Purcell like enhancement exists, and Umklapp processes have a significant contribution, see Figs.~\ref{fig:mat_3} and \ref{fig:mat_4}.

\FloatBarrier


